# Improving Accountability in Recommender Systems Research Through Reproducibility

**Alejandro Bellogín · Alan Said**

**Abstract** Reproducibility is a key requirement for scientific progress. It allows the reproduction of the works of others, and, as a consequence, to fully trust the reported claims and results. In this work, we argue that, by facilitating reproducibility of recommender systems experimentation, we indirectly address the issues of accountability and transparency in recommender systems research from the perspectives of practitioners, designers, and engineers aiming to assess the capabilities of published research works. These issues have become increasingly prevalent in recent literature. Reasons for this include societal movements around intelligent systems and artificial intelligence striving towards fair and objective use of human behavioral data (as in Machine Learning, Information Retrieval, or Human-Computer Interaction). Society has grown to expect explanations and transparency standards regarding the underlying algorithms making automated decisions for and around us.

This work surveys existing definitions of these concepts, and proposes a coherent terminology for recommender systems research, with the goal to connect reproducibility to accountability. We achieve this by introducing several guidelines and steps that lead to reproducible and, hence, accountable experimental workflows and research. We additionally analyze several instantiations of recommender system implementations available in the literature, and discuss the extent to which they fit in the introduced framework. With this work, we aim to shed light on this important problem, and facilitate progress in the field by increasing the accountability of research.

A. Bellogín
Universidad Autónoma de Madrid, Spain
E-mail: alejandro.bellogin@uam.es

A. Said
University of Gothenburg, Sweden
E-mail: alansaid@acm.org

## 1 Introduction

Evaluation of recommender systems is an active and open research topic. A significant share of recommender systems research is based on comparisons of recommendation algorithms' predictive accuracy: the better the evaluation scores (higher accuracy or lower predictive errors), the better the recommendation. However, given the lack of agreement on evaluation procedures, this poses a hurdle for reproducible environments. Considering this scenario, it is difficult to put in context and gauge the results from a given evaluation of a recommender system. One key reason for this is the existence of countless design alternatives when implementing an evaluation strategy. Additionally, the actual implementation of a recommendation algorithm can considerably diverge from the well-known, or ideal, formulation due to manual tuning and alignments to specific situations. This has been recently coined as data processing or data collection biases [65].

The most well-known recommender system-related event to date, the Netflix Prize, created an environment where all competing algorithms were evaluated in the same controlled environment. The research advances accomplished during the Prize's three-year run would not have been possible without a controlled evaluation and benchmarking environment. Having said this, we should also stress the importance of research conducted outside the scope of such actions. However, when it comes to recommendation, given that progress is traditionally measured in terms of higher precision, lower RMSE, higher NDCG, etc., it seems intuitive that some form of controlled evaluation could lead to a broader understanding of recommender system algorithms' qualities in general, and lessen the focus on in situ optimizations of deployment-dependent aspects.

In this context, reproducibility and replicability of experiments have been gaining the community's attention. Reproducibility is a cornerstone of scientific process and is one of two critical issues to achieve progress, as identified by Konstan and Adomavicious [53], and one of the priorities that should be reflected in experimental methodologies according to Ferro et al. [35]. However, as pointed out by Collberg and Proebsting [23], there are serious concerns in many research areas about this aspect. Collberg and Proebsting examined 601 conference and journal papers published by the Association for Computing Machinery (ACM). Among the papers, 402 had results backed by code, but the authors were able to obtain and compile the code within 30 minutes only for 32.3% papers. This opaqueness in research has led to directed efforts such as conference tracks aimed specifically for work covering these issues. Beyond those attempts at raising awareness on this issue, suggested solutions have been made in the community, e.g. Brinckman et al. [18] propose a general approach where research is presented as a living publication or tale. In doing so, improving and facilitating the knowledge discovery and dissemination of results, while supporting the entire life cycle of data used throughout research.

Synchronously, accountability and transparency of algorithms, or from a more general perspective, of any computational model, are gaining more atten-tion. For instance, Diakopoulos [27] and Lepri et al. [58] analyze accountability and transparency from the general viewpoint of algorithmic decision making. Both works state that these aspects can be considered at different levels: the entire model, individual components (e.g., parameters), and a particular training algorithm. On a similar vein as publications can be seen as living tales, Dragicevic et al. [30] suggest the use of explorable multiverses to increase the transparency of research papers, giving the reader the possibility to interactively explore alternative analysis options.

Going back to the Recommender Systems community, there exist few works addressing these aspects explicitly in the field. Although related areas, e.g. Human-computer Interaction or Information Retrieval, have pursued this line of work more actively. However, the notation and terminology is not consistent across research works, even within the same field. As an example, the work presented by Abdul et al. [2], which focuses on Human-computer Interaction, equates accountability to *explaining the decisions made by the algorithm*, which is more related to the concept of explainability; this definition differs from the one used in our work. One example where accountability is discussed in the context of recommender systems is the work by Ekstrand et al. [31]. The work argues that providing users with control over the generated recommendations would help establish trust to system, and lead to better retention and engagement. In this light, trust can be another dimension related, but not necessarily inherent, to accountability. Similarly, in the context of news Recommender Systems, some works have proposed implementations of transparent and "accountable" technologies to foster trust and use since, *there is an urgent need to explore how these technologies should be designed to effectively implement and promote data protection rights and obligations* [56].

Therefore, an important gap that remains is *how to develop accountable and transparent recommender systems in a general way*. In this work we, first, revise these aspects and their definitions for Recommender Systems research. Specifically, we present a brief review of different uses of these concepts in the general field of Artificial Intelligence, and later focusing on Recommender Systems. We then propose to address this issue by focusing on the reproducibility of the different stages that are typically defined within the workflow of any recommender system. Our driving hypothesis is that advances on reproducible environments for recommendation could provide better accountability of Recommender Systems research, understood as appropriate mechanisms for reporting evaluation results. For that, we propose reproducibility guidelines, inspired by how recommendation systems are implemented and evaluated, and argue that these should improve accountability of the entire research field. By doing so, researchers and practitioners would benefit from being able to reproduce and compare against other works, while the community would, in turn, gain as a whole. We should be able to increase our understanding of the recommendation components in use, explain why certain stages are more important in certain domains than in others, and, advance on the directions with more potential, thanks to well-grounded knowledge based on such accountable and transparent recommender systems. In the remainder of this section we present the research questions addressed in this work, as well as the main contributions achieved.

1.1 Research Questions

In this work we address the following research questions:

RQ1 **Which are the main aspects that obstruct accountability in Recommender Systems research?** In our work, we focus on how to increase transparency in reporting of experimental settings and corresponding results of recommender systems evaluation. Accountability is related to reproducibility, which is an increasingly popular topic in this area [32, 54, 70].

RQ2 **How can we formalize the different alternatives of the identified aspects in an accountable framework?** We propose a formal framework that embraces the different aspects related to accountability of recommender systems. For this, we start the analysis from lessons learned both from practical and theoretical perspectives in works addressing reproducibility in the area; we then analyze how different software libraries for the development of recommendation systems fit under the proposed framework.

RQ3 **What are the main challenges and limitations to improve accountability in Recommender Systems research?** Considering that accountability can be improved by focusing on reproducibility aspects of recommender systems, we study and present the main challenges and limitations that remain open in this scenario.

1.2 Contributions

We review concepts related to accountability and reproducibility, in particular in the context of Recommender Systems research, but also in adjacent contexts. Based on the definitions of those concepts, we propose a means to improve accountability of Recommender Systems research through reproducibility. This is achieved by analyzing technical constraints and typical problems or uncertainties that appear when implementing these systems. In particular, we have collected a wide range of lessons learned from practical, theoretical, and technical aspects about the implementation and evaluation of recommender systems. We have also considered recent works and trends on reproducibility in neighboring fields to extract lessons and guidelines for recommender systems.

Upon this, we propose a set of requirements that lead to more reproducible experimental workflows, improving the mechanisms necessary for attaining accountable Recommender Systems. According to these requirements, we introduce a framework that consists of six components: dataset collection, data splitting, recommendation, candidate item filtering, evaluation computation, and statistical testing. Moreover, we analyze to what extent current Recommender Systems libraries fit such framework.

Finally, we also include a thoughtful discussion on the limitations of this proposal, complemented with the foreseen challenges that need to be addressed in the near future. This discussion is complemented with experiments where we compare results from the internal evaluation mechanisms of two popular open source recommendation frameworks. By doing so, we show that a lack of control on the recommendation process leads to inconsistent results and, hence, to non comparable and reproducible experimental outcomes. We believe the general framework and the presented analyses and discussions will be valuable for practitioners, designers, or engineers who aim to assess the capabilities of published research on recommender systems.

1.3 Structure of Paper

The rest of this paper is structured as follows. Section 2 presents in detail the concepts that will be considered in this work (accountability, transparency, reproducibility) and provides an overview of recommender systems, how they are evaluated, and implemented. Section 3 discusses our approach of facilitating accountability through reproducibility, for this, prior works related to reproducibility in evaluation are discussed together with several technical aspects that should be considered when implementing and evaluating a recommender system. Then, in Section 4, we introduce the requirements for a reproducible and accountable recommendation environment, followed by an analysis of how the most common software libraries satisfy these requirements. Finally, Section 5 analyzes the main limitations and challenges of the presented work, and Section 6 concludes the paper and presents future work directions.

**2 Background and concepts**

In this section, we provide an overview of the two foci of this work: accountability and recommender systems. The latter is presented and contextualized in Section 2.1, whereas the former, together with other related concepts used throughout this work, is introduced in Section 2.2.

2.1 Recommender systems

Recommender systems have become ubiquitous means of assisting users' in online and offline services, for music (Spotify, Apple Music), movies and videos (Netflix, YouTube, Hulu), and for retrieval of more heterogeneous items (Amazon, eBay, Facebook, etc.). With the growing popularity of recommender systems in recent years, the recommender system-related research field has grown dramatically. Today, paper tracks on recommendation are common in many research outlets which traditionally have had different foci, e.g. SIGIR, The Web Conference, IJCAI, etc. Related industry has experienced a similar development. As a sign of this, many open positions in machine learning, data science, and artificial intelligence list experience of recommendation techniques and frameworks as requirements.

This gain in popularity has led to an overwhelming amount of research being conducted over the last years. With this in mind, it becomes increasingly important to be able to reproduce new approaches and gauge their performance in various settings. In today's recommender systems-related literature, papers often state what datasets, algorithms, baselines, and other potential parameter values are used in order to (theoretically) ensure reproducibility. Many research manuscripts additionally provide access to source code, present execution times, necessary hardware and software infrastructure, etc. Nonetheless, the works often use proprietary data and software making it difficult, or impossible, to accurately reproduce the findings. In light of the vast variety of methods available for implementing and evaluating a recommender system, details which aid in reproducibility and transparency should be seen as a very positive aspect. A critical analysis is necessary in order to ensure an advance in the field, not just marginal effects based on strategic design choices [33].

In fact, there have, over time, emerged several popular recommendation frameworks used throughout academia and industry, e.g. *Apache Mahout* [1] (henceforth referred to as Mahout) [66], *LensKit* [2] [32], *MyMediaLite*[3] [39], *LibRec*[4] [43], and *RankSys*[5] [80], among others. As common with open source frameworks, popularity and development efforts fluctuate over time. This can be seen in, for instance, MyMediaLite which was very popular a number of years ago, although its popularity dwindled as development came to a standstill. Mahout was once very popular in the recommender systems community, but as the project changed direction and focused development on other tasks, the library has become seldom seen in research works in this space today. Nonetheless, we chose to include these two frameworks in the analysis presented in this paper due to their historical popularity and extensive usage in research publications and industry applications. Moreover, even though all five frameworks listed above provide basically the same set of recommendation algorithms, they have various differences in implementations, data management, and evaluation methods and approaches. Moreover, each of these frameworks provide basic evaluation packages able to calculate some of the most common metrics (e.g. precision, recall, root mean square error, mean average error, etc.), however, it is not clear how they compare against each other. Later, in Section 4.2 we will analyze the most representative ones and compare their differences.

In any case, there is no doubt that these, and other, open source recommendation frameworks, have been and are used consistently in research. However, the lack of a standard framework or implementation of algorithms and evaluation methodologies, impedes the progress in the field, as evidenced in other areas. In this context, there are examples showing that, in e.g. the Information Retrieval area, even when standard datasets are used and a well-established set of baselines is known, no overall improvement over the years is guaranteed [5]. A similar example exists in Ferrari Dacrema et al.'s [33] analysis of deep learning approaches for recommender systems, and in Sun et al.'s recent horizontal study of recommender systems papers [79]. Analogously, in a series of prior works focusing on the evaluation, replication, and reproducibility of recommender systems algorithms and evaluation results, we have identified a set of aspects that need to be taken into consideration when comparing the results of recommender systems from different research papers, software frameworks, or evaluation contexts [70, 72, 71]. Specifically, a previously performed study [70] pointed out that comparisons of the same algorithms and datasets across different implementations (i.e., recommendation frameworks) showed large variations in results, both in accuracy-based metrics (like precision) and in holistic metrics (such as catalog coverage).

---

[1] http://mahout.apache.org
[2] http://lenskit.grouplens.org
[3] http://mymedialite.net
[4] https://guoguibing.github.io/librec/index.html
[5] http://ranksys.org

2.2 Accountability, Reproducibility, and Related Concepts

Researchers from different areas dealing with algorithms – from Social Sciences to Data Mining and Machine Learning, including the Information Retrieval and Recommender Systems communities – have been raising awareness about accountability, reproducibility, transparency, fairness, explanation, and similar concepts in recent years. However, these terms may have different nuances depending on the context, or might even appear as synonyms according to some authors. To establish a common terminology for this work, this section presents a review of such definitions and states how they will be used in the rest of the manuscript.

Starting with reproducibility and replicability, we want to bring attention to the fact that these terms are often confused. In Plesser's work on confused terminology [67], the author presents a historical review of the terms, discussing differences and similarities to previous works. In this paper, we follow the definitions provided by the ACM[6]: **repeatability** applies when the same team operates under the same experimental setup, **replicability** is achieved when a different team obtains the same results using the same experimental setting, and **reproducibility** is obtained when a different group using their own experimental setup produces the same results. In fact, Plesser points out that the definitions of replicability and reproducibility used by the ACM, while conforming to their definitions in computing, are at odds with the terminology long used in the experimental sciences (we refer the reader to [67] for more details about these works). Ivie and Thain [45] discuss to what extent scientific computing is accurately reproducible. They extend the terminology around these concepts with terms such as verification, validation, and provenance, to account for variations in the data or tasks that could apply to computational experiments. The work presents several technical challenges that may appear when replicating different artifacts (commands or workflows). It does also show general techniques that could help in achieving reproducibility, mostly related to tracking all operations and actions related to an experiment and giving more importance to validation.

---

[6] https://www.acm.org/publications/policies/artifact-review-badging

Continuing, **accountability** is defined by Lepri et al. [58] as the assumption of accepting the responsibility for actions and decisions. Moreover, an ACM statement on Algorithmic Transparency and Accountability[7] calls for better algorithms in a number of dimensions since *the ubiquity of algorithms in our everyday lives is an important reason to focus on addressing challenges associated with the design and technical aspects of algorithms and preventing bias from the onset*. The statement lists the following 7 principles: awareness, access and redress, accountability, explanation, data provenance, auditability, validation and testing. In particular, accountability is described as "Institutions should be held responsible for decisions made by the algorithms that they use, even if it is not feasible to explain in detail how the algorithms produce their results". According to Jobin et al. [49], very different actors are pointed out as being responsible (and accountable for) the actions and decisions of an Artificial Intelligence agent: developers, designers, institutions, or industry at large. In particular, Shin et al. [76] specify that accountability in algorithms and their application begins with the designers and developers of the system that relies on them. Eventually, designers and managers are responsible for the consequences or impacts an algorithmic system has on stakeholders and society. However, there is no consensus on whether these agents should be held accountable in a human-like manner or whether humans should always be the only actors who are ultimately responsible for technological artifacts [49].

A related concept is that of **transparency**, which Lepri et al. [58] understand as the ability to provide openness and communication of both the data being analyzed and the mechanisms underlying the models. Going beyond this definition, certain works discuss the implications of the concept. For instance, Shin and Park [76] observe that perceptions of transparency and accuracy are not purely objective responses to media content. In this sense, transparency and accuracy cannot be measured as features of algorithms or in the context of legal terminology because they involve subjective dimensions that can be experienced and perceived by users. Diakopoulos [27] takes a more general perspective and argues that transparency can be a mechanism that facilitates accountability. Hereto we should demand it from government and exhort from industry, however, requiring complete source-code transparency of algorithms may be counterproductive. Instead, Diakopoulos advocates for the disclosure of only necessary pieces of information, including aggregated results and benchmarks, since these elements would be more effective in communicating algorithmic performance to the public.

---

[7] https://www.acm.org/articles/bulletins/2017/january/usacm-statement-algorithmic-accountability

Moreover, transparency around human involvement might involve explaining the goal, purpose, and intent of the algorithm, including editorial goals and the human editorial process. More specifically, in research the control is on the researchers, or the authors of an investigation, hence, according to this definition, they have to disclose as much information as possible to provide transparency and improve its accountability. The same author adds that inferences made by an algorithm, such as classifications or predictions, often leave questions about the accuracy or potential for error. Because of that, algorithm creators might consider benchmarking against standard datasets and with standard measures of accuracy to disclose some key statistics. This is directly connected to the concepts of reproducibility and replicability mentioned earlier.

Finally, to conclude this section, we survey works where these concepts have been explicitly applied to the Recommender Systems area. At least since 2005, these topics have gathered interest from the community. We find the following statement by O'Donovan and Smyth [64] about trust in recommender systems: *Our trust models can be used as part of a more broad recommendation explanation. (...) We believe that this recommendation accountability is a very influential factor in increasing the faith a user places in the recommendation.* Here, there is a belief that adding explanations (as one possible tool to provide accountability of the system's actions) would increase the trust of the user in the system. More recently, Ekstrand et al. [31] discuss accountability in the context of recommender systems under a similar view, where the authors argue that providing users with control over the generated recommendations would help users trust the system and lead to better retention and engagement. From a different viewpoint, Jobin et al. [63] equate good reproducible standards with accountable recommender systems. They propose to permanently store and publicly display the configuration parameters and numerical outcomes of the algorithms involved in an experiment to make offline comparisons easier.

The number of works explicitly dealing with reproducibility and replication issues for recommender systems has increased in the last years. This is partially due to the existence of dedicated reproducibility tracks, as we will discuss in Section 3.2. Beyond the papers previously mentioned [70, 79, 33], other works have explored these issues from novel perspectives. For instance, Beel et al. [9] analyze reproducibility in the news domain and research-paper recommendation. Since these domains are so different to each other, the authors found that there are large discrepancies between them even when the same approach is used. An important outcome of this analysis is that the authors obtained several *determinants* that may contribute to these discrepancies: user characteristics, the time at which recommendations were shown, the user-model size, and so on. Coba and Zanker [22], on the other hand, performed a usability study on an open-source library and found that users appreciated the transparency provided by such tools.

# 3 Accountability of Recommender Systems Research through Reproducibility

In this section, we motivate how accountability (as defined in the previous section) could be achieved in Recommender Systems research by means of reproducibility. First, in Section 3.1 we introduce some lessons learned from practical aspects found when implementing these systems, later in Section 3.2 we provide a more general overview of trends and research works that have explicitly addressed reproducibility in the area. Finally, Section 3.3 summarizes and connects these lessons and trends while presenting how they must be exploited to improve accountability.

## 3.1 Reproducibility Lessons from Practical and Technical Aspects in Recommender Systems

There are several aspects when designing and implementing a recommender system that may affect its final results and hinder its transparency, making it virtually impossible to replicate the reported settings and insights. From using different model parameters to how specific parts of the system are implemented. In this section, we review some examples we have encountered while implementing recommendation algorithms and frameworks ourselves, checking public implementations, and trying to replicate other researchers' experiments and results [70, 71]. In the next section, we propose the requirements that any evaluation framework should satisfy to facilitate reproducibility of the research works, based on these examples.

One of the first steps taken when running a recommender system consists of dividing a dataset into training and test splits, in some circumstances a validation split is also created [75]. It is understood that the data splitting strategy may have a deep impact on the final results, making this an important aspect to take into account when reporting details about an experiment. For instance, by using a per-user data splitting technique, we ensure that all the users are considered in the evaluation. This may not be realistic in all cases, since all users are not active simultaneously. On the other hand, by performing a temporal split – which likely is the closest data splitting strategy to a real world scenario [19] – the models are more sensitive to temporal dependencies on the training and test splits. For instance, temporal spikes on particular items that might skew popularity patterns and not represent the historical behavior in that system. In both cases, the data splitting strategy dramatically changes the inherent conditions of the experiment. Whereas in the first case, every user is active at evaluation time. In the second case, the set of active users will depend on the level of interaction at, and after, the temporal point where the split is performed, which, additionally, ensures no future information is used to train the models.

Nonetheless, most of the practical issues related to reproducibility concern either technical implementations of the actual algorithm, or how the evaluation of the system was performed. We now review several examples regarding these two angles.

Regarding the **recommendation algorithms**, there are several aspects that are not standardized which, eventually, turn out to be implementation-dependent. For example, the term kNN (to specify: collaborative filtering algorithms based on k-nearest neighbors) is usually linked to the user-based (UB) variation of these methods. However, item-based (IB) also fall into the kNN category, and often outperform the user-based counterparts [74]. Furthermore, there exist several different formulations of the user-based kNN algorithm, neither of which is regarded as the "standard" one. It could be argued that Goldberg et al.'s [40] initial formulation can be seen as the *classical* standard formulation, and that other formulations are simply variations that aim at achieving better performance. In this context, each of the formulations in Equations 1, 2, or 3 could be referred to as UB kNN, and unless further details are provided, knowing which of the three is used by the authors of a paper is virtually impossible:

$$\hat{r}_{ui} = \frac{\sum_{v \in N_f(u)} r_{vi} w_{uv}}{\sum_{v \in N_f(u)} |w_{uv}|} \quad (1)$$

$$\hat{r}_{ui} = \bar{r}_u + \frac{\sum_{v \in N_f(u)} (r_{vi} - \bar{r}_v) w_{uv}}{\sum_{v \in N_f(u)} |w_{uv}|} \quad (2)$$

$$\hat{r}_{ui} = \sum_{v \in N_f(u)} r_{vi} w_{uv} \quad (3)$$

Note that in one of these formulations (Equation 2) the user's average rating $\bar{r}_u$ is considered, whereas in the last case the summation is not normalized, with the effect that the generated score is not in the range of ratings anymore. This variation has proved to obtain much better results than the others in tasks that return a ranking instead of predicting the rating [24]. In fact, some libraries implement two versions of these methods (such as MyMediaLite or LibRec), highlighting the importance of providing these details.

Analogous to algorithmic variation, the similarity weight between two users (denoted $w_{uv}$ above) is a critical aspect for kNN recommenders, and decisions made about them can profoundly affect how the algorithm behaves. The decision on which similarity metric to use should be clearly specified and reported. Then, any filtering or transformation applied to the similarity values must be clearly specified. For example, some libraries like Mahout[8] move the similarity domain from $[-1, 1]$ to $[0, 2]$ to avoid negative weights in the summation. Other implementations simply discard those negative similarities, or allow to discard anything below a specific threshold (see LibRec[9] or MyMediaLite[10]). In other cases, two similarity metrics are used: one to find the neighbors (i.e., to define $N(u)$) and a different one to weight the ratings in whatever variation of the UB kNN algorithm we are using – although it is more common to use the same metric for both steps; the same comments about similarity transformation and filtering apply if it is used for finding neighbors.

---

[8] See normalizeWeightResult function in https://github.com/apache/mahout/blob/trunk/community/mahout-mr/mr/src/main/java/org/apache/mahout/cf/taste/impl/similarity/AbstractSimilarity.java (retrieved: April 2020).

There also exist different alternatives in the literature regarding the task of finding neighbors. Although this typically means taking the top-k closest users to the target user, there are other techniques based on thresholds that should be properly reported and specified when used in experiments because of their unpopularity [25]. There are also ad hoc implementations in some libraries[11]. Besides that, a very important detail that is often overlooked is the exact moment at which neighbors are found: at training or test (i.e., evaluation) time. In both cases, similarities between users are needed, however in the latter approach the computation of the list of potential neighbors is delayed until a prediction is needed. This means that the neighborhood can be tailored to those neighbors that have actually rated the target item (as evidenced in the notation used before as $N_i(u)$ instead of simply $N(u)$), achieving higher values of coverage and impacting on the optimal number of neighbors required (lower for the delayed computation) – this can be found in Surprise[12], for instance. Nonetheless, this approach would not be practical in a real world system, as it would require having instantaneous access to either the full similarity matrix, or the full list of users sorted by similarity for every user. Both issues are avoided by computing the full similarity matrix offline, and only storing the top-k closest neighbors to be used at test time.

Another paradigmatic example, besides the different formulations for kNN and their related issues presented above, is how a user's average rating is computed (a parameter needed by some similarity metrics, such as Pearson's correlation, and in Equation 2). In theory, the user's average rating should be computed taking into account all her ratings, however this implies a first pass on the whole dataset is needed and a value for every user is stored in memory. For computational reasons, it is reasonable to delay this process until a prediction involving the user is needed. The issue becomes evident when, in some implementations, this computation is only performed for the items that a user pair have in common (for example, in Mahout[13]). This would generate noticeably divergent results for different implementations, even though the reported algorithm is the same; see Table 1 for an actual comparison. In this toy example, we observe how the similarity score between users changes depending on how the users' average ratings are computed; although the actual difference is small (from 0.96 to 1.0) it illustrates the potential divergences that could accumulate with more data.

---

[9] See predict function in https://github.com/guoguibing/librec/blob/3.0.0/core/src/main/java/net/librec/recommender/cf/UserKNNRecommender.java (retrieved: April 2020).

[10] See GetPositivelyCorrelatedEntities function in https://github.com/zenogantner/MyMediaLite/blob/master/src/MyMediaLite/Correlation/Extensions.cs (retrieved: April 2020), which is used in the UserKNN and ItemKNN implementations.

[11] See note about SnapshotNeighborhoodFinder in https://java.lenskit.org/documentation/algorithms/user-user/ (retrieved: April 2020).

[12] See estimate function in https://github.com/NicolasHug/Surprise/blob/master/surprise/prediction_algorithms/knns.py (retrieved: April 2020).

[13] See functions userSimilarity and itemSimilarity in https://github.com/apache/mahout/blob/trunk/community/mahout-mr/mr/src/main/java/org/apache/mahout/cf/taste/impl/similarity/AbstractSimilarity.java (retrieved: April 2020).

**Table 1** Toy example showing the value of a user's average rating considering all items ($\bar{r}_u$) or only those in common with the other user. Pearson's correlation similarity metric between users ($w_{uv} = \sum_i (r_{ui} - \bar{r}_u)(r_{vi} - \bar{r}_v) / \sqrt{\sum_i (r_{ui} - \bar{r}_u)^2 \sum_i (r_{vi} - \bar{r}_v)^2}$) is also presented when considering these two variations for the average rating.

| Items | User 1 | User 2 |
|---|---|---|
| $i_1$ | 5 | 3 |
| $i_2$ | 3 | 1 |
| $i_3$ | 4 | 2 |
| $i_4$ | 4 | |
| $i_5$ | | 3 |
| $\bar{r}_u$ | 4 | 2.25 |
| Overlap $\bar{r}_u$ | 4 | 2 |
| Pearson using $\bar{r}_u$ | | 0.96 |
| Pearson using overlap $\bar{r}_u$ | | 1.0 |

**Table 2** Toy example showing different values of Mean Absolute Error (MAE) and Root Mean Squared Error (RMSE) values depending on how missing predicted values (denoted as NaN) are considered by the metric. The best system in each case is highlighted in bold.

| User-item pairs | Real | System 1 | System 2 | System 3 |
|---|---|---|---|---|
| $(u_1, i_1)$ | 5 | 4 | NaN | 4 |
| $(u_1, i_2)$ | 3 | 2 | 4 | NaN |
| $(u_1, i_3)$ | 1 | 1 | NaN | 1 |
| $(u_2, i_1)$ | 3 | 2 | 4 | NaN |
| MAE/RMSE, ignoring NaNs | | 0.75 / 0.87 | 2.00 / 2.00 | **0.50 / 0.70** |
| MAE/RMSE, NaNs as 0 | | **0.75 / 0.87** | 2.00 / 2.65 | 1.75 / 2.18 |
| MAE/RMSE, NaNs as 3 | | 0.75 / 0.87 | 1.50 / 1.58 | **0.25 / 0.50** |

The last aspect about recommendation algorithms we focus on is related to the final predicted score. Until relatively recently, much of the recommendation literature focused on the rating prediction task, overlooking other tasks, and, particularly the most prominent one nowadays: item ranking. As a result, it is easy to find implementations where a capping is applied after the score is predicted, binding the score by the rating range. However, this may be problematic when producing a ranking instead of a rating. Since ties are more likely to occur, leading to the necessity of tie-breaking strategies. These may, however, have strong effects on the results, especially when short rankings are considered [10]. This is something seldom found explicitly in the analyzed frameworks and public implementations[14].

---

[14] In particular, among the analyzed frameworks, only RankSys sorts the items according to anything else besides the score.

Furthermore, what is even more important is what happens when a recommender cannot predict a score. This decision has an impact on the system's coverage. If a baseline (also called *backup* or *fallback*) recommender is used, all algorithms will have full coverage. This will, in turn, also have an effect on how performance metrics are evaluated. For instance, when producing a ranking, one may decide that those items without a score will always appear at the bottom of the ranking. This could produce very different results compared to skipping those items altogether, especially in cases with extreme coverage problems or when using highly configurable evaluation strategies (such as the One-Plus-Random methodology from [11]). Additionally, when computing error-based metrics such as Mean Absolute Error, which computes the difference between the predicted and the actual rating, handling of missing values is not evident: they could be ignored (hence measuring the error only on available predictions) or a default value could be used instead (as if a baseline recommender would be used); see Table 2 for a toy example on these different alternatives. In the analyzed frameworks, various solutions have been detected: using a recommender that can always return a score whenever the main one is not capable of that[15], using default values when some operations (such as similarities) do not have enough data or cannot be computed[9], or leaving the original recommender untouched, hence producing missing values and exposing its lack of coverage.

Now let us move to the issues related to **how the system is evaluated**. As presented before, technical decisions about the algorithm implementation could affect the results of evaluation metrics. Moreover, as mentioned earlier regarding the algorithms, in recommender systems there are no well-established definitions for all the evaluation metrics (yet), as is common in other areas like Machine Learning or Information Retrieval. Thus, it is possible to find papers presenting different variations of the Mean Absolute Error metric (either the classical one, or normalized by the range of ratings, or even computed on a user-basis and then averaged), cutoffs for ranking metrics not being reported, and implementations where it is not clear which discounting function is being used for the Normalized Discounted Cumulative Gain (NDCG) metric or how the relevance score is obtained from the rating withheld in test [48][16].

Another important aspect is that of the candidate item selection. As studied by Bellogín et al. [11], the strategy used to select which items are to be ranked by the recommender plays a significant role in both the range of the performance values and the relative order of the recommendation algorithms. While this fact is being increasingly recognized by the community [57], there are still several gaps regarding this critical part of the recommen-dation pipeline, the most obvious one consists of acknowledging which strategy is being used and report it accordingly, along with any parameter required to accurately reproduce such experimental setting.

---

[15] As described in https://java.lenskit.org/documentation/basics/configuration/ (retrieved: April 2020), LensKit provides *default settings* which should be overriden to have full control; for an example of how MyMediaLite predicts a score by using a default recommender see https://github.com/zenogantner/MyMediaLite/blob/master/src/MyMediaLite/RatingPrediction/UserKNN.cs (retrieved: April 2020).

[16] Specifically, we have found that LensKit, LibRec, Mahout, MyMediaLite, and RankSys use a logarithmic discount, although RankSys normalizes the relevance with respect to a provided threshold and others (like Mahout) do not consider the rating in the computation.

3.2 Reproducibility Lessons from other Research Works and Recent Trends

As mentioned before, reproducibility in Recommender Systems research focuses on how reproducible the evaluation of such systems really is. This is evidenced in, e.g., the call for papers of the reproducibility track[17] of the 2020 ACM Recommender Systems conference (RecSys), where papers should *repeat prior experiments with original source code using original datasets or new contexts*. However, in, e.g. Information Retrieval, there is a growing concern that reproducibility is a process which in particular calls for the long-term sustainability of reproducibility efforts, as supported by Lin and Zhang's work [60], where the authors were not able to replicate a number of reproducible experiments four years after they had been published. Also coming from the Information Retrieval community, we find recent works dealing with the problem of generality of results, as well as *negative results* and design choices when implementing tools and evaluation campaigns with this goal in mind [4, 36, 37, 17]. Lessons learned from these works include e.g., sometimes it is not obvious how a (theoretical) method really is implemented in a framework, in particular due to potentially hidden and undocumented parameters, or that the implementation might change between versions of the same software [4]. Breuer et al. [17] analyze the important question of *when reproduced is reproduced*, that is, how to objectively quantify that an experiment has been replicated or reproduced. In this context, the work proposes several measures at different granularities and introduced a reproducibility-oriented dataset for Information Retrieval tasks.

Even though reproducibility tracks have become a common component in academic conferences in recent years, many of the topically-relevant conferences still lack one. One of the first tracks of this type was at a workshop at ACM Special Interest Group on Data Communication (SIGCOMM) in 2017 [1]. Also in 2017 a series of workshops on this topic started in the Machine Learning community, followed by the 2017 and 2018 editions of the International Conference on Machine Learning (ICML) and the 2019 International Conference on Learning Representations (ICLR). ACM RecSys had its first installment of the reproducibility track in 2020. Additionally, at the time of writing, we found the 2019 and 2020 editions of the European Conference on Information Retrieval (ECIR), ACM MultiMedia (MM) in 2020 – this was the first time in this conference and only accepted authors with an accepted paper in the previous edition of the conference, since the goal is to document and record the artifacts needed to carry out the experiments of the previously published research work –, and the 2020 edition of the International Seman-tic Web Conference (ISWC) – again, only for accepted authors but, in this case, of the same edition. A similar idea is behind the recent Reproducibility Challenge at ICLR. While these tracks are relatively new, similar initiatives have been surfacing during the last decade, e.g. as a workshop at ACM RecSys in 2013 [13]. These venues allow researchers to analyze the effect of different implementations of an approach, or explore the extent to which the results may change when a different dataset than the one reported in a paper is used [55, 50]. Often, the works published in these tracks contain publicly available code and/or datasets. However, a common complaint on these papers is the difficulty of perfectly reproducing the experiments from the papers, even when the code is available (which is not very common [23]), and in some cases the attempts have to be discarded because inquiries sent to the original authors related to code or data remain unanswered [33, 60, 50].

---

[17] Full text in https://recsys.acm.org/recsys20/call/#content-tab-1-1-tab (retrieved: April 2020).

A number of conferences have adopted ACM badges[6] and award these according to different criteria: functional, reusable, available, validated, reproduced, replicated. However, at the time of writing, all application procedures for these badges are not established yet. In general, badges are awarded to works that already meet some established standards (i.e., the paper is already accepted to a conference or journal). An instantiation of this process can be found for the MM conference[18], where, as a significant difference in the standard review process, reviewers of the reproducible papers become co-authors of the final reproducibility paper due to the expected heavy burden necessary to review such submission. This can be seen as hard evidence of the work load required to make sure that a research work is reproducible.

Finally, it is worth mentioning the report from a recent Dagshtul seminar dedicated to the reproducibility of data-oriented experiments in e-science [38], even though the Recommender Systems community is omitted in the report. The report mentions that reproducible experiments are not only beneficial for others, but also for the researcher or practitioner executing them. The reasoning being that they require the researcher to document the entire experimental settings, which makes post hoc analysis possible. As a result, the report states that the experiment can be easily audited, making accountability quite straightforward. The report presents tools that may improve the reproducibility of the experiments, and additionally, the output of several working groups which addressed different aspects of the reproducibility problem: who are the main actors, defining guidelines for authors, editors, and others, and a list of the incentives and barriers to reproducibility.

### 3.3 Summary and discussion

Based on the previous definition of accountability, we propose that the label *accountable Recommender Systems research* is given to works dealing with recommender systems while the executing parties (the researchers) are held accountable for the presented results, and to some extent (by considering the extended definition of the ACM statement on Algorithmic Transparency and Accountability), when these experiments are explainable by the researcher, in the sense that the results can be generated in a deterministic and guided way, making them possible to be audited.

---

[18] See https://project.inria.fr/acmmmreproducibility/ (retrieved: April 2020).

In this section, we have collected a series of lessons learned either from practical and direct connection with the implementation of recommender systems, or from published papers and other venues where reproducibility has been studied by itself. We argue that, by increasing the reproducibility of research works, accountability will improve, since reproducible environments guarantee that the results can be audited and, under some constraints, repeated, facilitating that the same conclusions will be found. Moreover, in recent years, the opposite seems, unfortunately, to be true. Recent studies indicate that the results of papers which are difficult to reproduce, generate doubts about their conclusions and, in turn, the responsibility of the authors for the actions or decisions drawn in the paper will not hold [23, 33].

Therefore, we advocate for a reproducible environment in Recommender Systems research and, in particular, for evaluation of these types of systems, so as to enforce certain levels of accountability in this ever-growing research field. In the next section, based on the lessons learned and conclusions drawn in previous sections, we propose a set of requirements as potential guidelines to achieve reproducible and accountable recommender system evaluation.

## 4 A Reproducible (and Accountable) Recommender Systems Framework for Evaluation

Building on experiences described in the previous section, a study on recommender system evaluation across software frameworks [70], works on benchmarking and evaluation [72, 11], as well as on reproducibility of evaluation results [71], a set of requirements for ensuring the accountability and reproducibility of recommender systems have been identified as necessary tasks in the recommendation process. This section gives, first, an overview and motivation of these tasks, and later a possible instantiation of such a framework, linking it to the available libraries in the Recommender System area.

### 4.1 Requirements for an Accountable Recommender System Evaluation

Figure 1 presents a step-by-step overview of the research process in recommender systems. The various stages in the figure, i.e., *Dataset collection*, *Data splitting*, *Recommendation*, *Candidate Item Filtering*, *Evaluation*, and *Statistical Testing*, correspond to separate stages which form the complete pipeline of a typical research recommender system; next, we describe each of these stages.

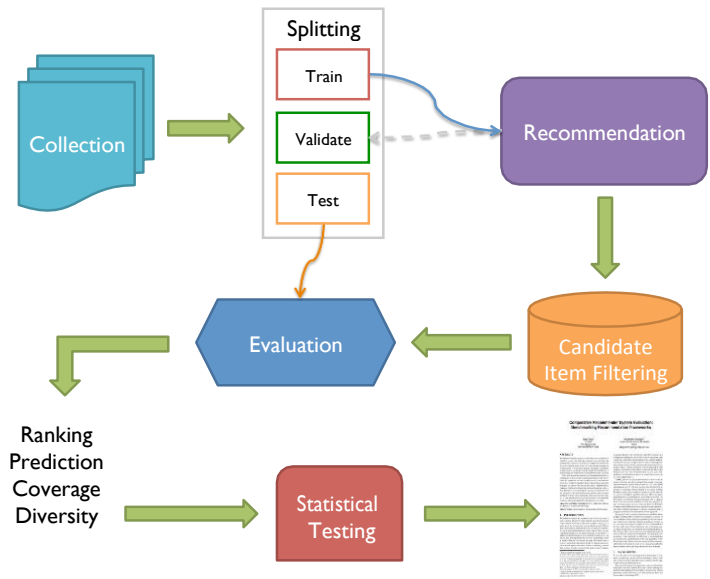

**Fig. 1** The recommendation process distilled into several sub tasks or stages.

#### 4.1.1 Dataset Collection

The collection step represents the initial stage where data is collected. Documenting the process of how the data was collected, and whether, e.g., certain prefiltering or other modifications were performed on the data. This is often overlooked in papers, especially so in cases where public datasets are used. Nevertheless, it is important to consider how the dataset has been altered prior to having been released to the public. Examples of this include Movielens20M[19], where it is stated that "All selected users had rated at least 20 movies", the 2017 RecSys Challenge [3] where the dataset preprocessing details are withheld, and MovieTweetings [29] which states that the data is collected via a third party which in turn does not guarantee completeness.

#### 4.1.2 Data Splitting

The data splitting procedure, taking place before recommendation, is an important step in the experimental configuration of a recommender system. How datasets are partitioned into training and test sets may have considerable impact on the final performance. It may be favorable to certain recommenders, and hurtful to others, consequently resulting in better or worse performance depending on how the partitioning is configured. A few common approaches for data splitting follow below.

First, we can choose whether or not splitting will be temporal [42]. Temporal approaches require the availability of user interaction data timestamps.

---

[19] http://files.grouplens.org/datasets/movielens/ml-20m-README.html

A simple approach is to select a point in time in the available data timeline, and separate training data (all interaction records prior to that point) and test data (dated after the split time point). The split point can be set so as to, for instance, have a desired training/test ratio in the experiment. The ratio can be global, with a single common split point for all users, or user-specific, to ensure the same ratio per user. Temporal approaches have the advantage of more realistically mimicking working application scenarios, where "future" user positive interactions (which would translate to positive response to recommendations by the system) are to be predicted based on past evidence. As an example, the well-known Netflix prize provided a dataset where the test set for each user consisted on her most recent ratings [14].

There are a number of common strategies ignoring time. We outline the following three strategies that select the items to appear in the test set: *a)* sample a fixed number (potentially different) for each user – this is a common approach to test cold-start scenarios, as done by e.g. Kluver and Konstan [51]; *b)* sample a fixed number for each user (equivalent for all users), also known as *given n* or *all but n* protocols; *c)* sample a percentage of all the interactions using *cross validation*, where non-overlapping subsets are generated, that is, every (user, item, rating) tuple is evaluated once. The last protocol, *c*, appears to be the most popular [41, 74], although *all but n* protocol is not uncommon [16, 81].

Independently from the dataset partitioning, the goals set for an evaluation may be different depending on situation, requiring the development of different settings (and, in turn, partitioning protocols) [44, 42]. If this is ignored, the results obtained in a particular setting can become biased, and thus not suitable to use in further experiments. In light of this, specifying which partitioning protocol is used in the experiments is a crucial parameter to ensure reproducibility and objective results.

### 4.1.3 Recommendation

Once the training and test data partitions are available, recommendations are produced according to different algorithms. In an evaluation, the recommendation component is the only one that should differ when comparing two recommendation algorithms. In particular, one and the same algorithm implemented in two different recommendation frameworks should produce equivalent recommendations when fed with the same input. However, after performing a side-by-side comparison of the same algorithms in different frameworks, we see that the recommendations can differ widely, as reported in previous work [70]. There are multiple reasons for this, some of them are listed in the previous section, such as design choices related to the implementations and different algorithm parameter settings.

### 4.1.4 Candidate Item Filtering

Candidate item filtering, or *candidate item generation* as it sometimes is referred to, specifies which of the recommended items will be used to measure the performance of the recommendation, in particular, the items that actually generate a ranking. It may be reasonable to perform this step prior to recommendation, however, the recommendation process described in Figure 1 assumes that the previous stage predicts all the available recommendations for a user, and hence, in this step a filtering is performed. In the event of application of multiple strategies, this ordering of the steps is more efficient, since it allows to compute the recommendations only once.

Moreover, we may want to measure recommendation performance based on predictive accuracy (the error with respect to the predicted rating) or ranking-based precision. Both gauges have their limitations: whereas the former needs the recommender to be able to output a score meaningful as a rating, the latter is sensitive to the number of unknown items included in the generated ranking, and how they are selected [11]. We shortly describe two strategies below (we refer to prior works on this topic, e.g., [70, 11] for more).

UserTest This strategy takes the same target item sets as error-based evaluation. That is, for each user $u$, the list $L_u$ of candidate items to be ranked consists of items rated by $u$ in the test set. This strategy selects the smallest set of target items for each user, including no unrated items at all. It needs a relevance threshold to indicate which of the items in the user's test are considered relevant. This strategy is used by, e.g. Jambor and Wang [46, 47] with the relevance threshold set to 4 in a 1-5 rating scale, and by Basu et al. [8] with an individual, per-user, threshold.

TrainItems Every item rated by some user in the system is selected – except items rated by the target user. This strategy is useful when simulating a real system where no test is available (at prediction time) because it does not need to look into it to generate the rankings.

### 4.1.5 Evaluation

This stage is devoted to the task of measuring the performance of the produced scores, i.e. to evaluate them. Depending on the type of evaluation metric, the previous step described (candidate item filtering) may or may not be required. Specifically, for error-based metrics like mean absolute error (MAE) and root mean square error (RMSE), predicting only for the items in the test set is adequate, even though in the recommendation step (Section 4.1.3) predictions for all the items in the dataset might be produced. Nevertheless, reporting coverage of the methods is essential – both in terms of user coverage (number of users that receive at least one recommendation) and catalog coverage (percentage of all the items that can be recommended) [75]. Considering, that if we do not use a baseline or *backup* recommender, the recommender might not return all requested suggestions. If we expect to receive 10 recommendations for a user but only get 8 from a specific recommender, we would measure the performance based on those 8 items instead of the requested recommendation length. As stated by Herlocker et al. [44], this is a well-known problem, the only solution to date, is to report both accuracy and coverage metrics at the same time, although some approaches have been proposed recently to address this issue [62].

Ranking-based metrics like precision, recall, normalized discounted cumulative gain (NDCG) and mean reciprocal rank (MRR) aim to capture the quality of a particular ranking, taking into account that the user has expressed a preference towards some of those items, typically those in the test split rated above a certain threshold [6]. The candidate item filtering stage will change the value computed with these metrics. This is due to a different ranking being evaluated, although the order of the recommenders does not necessarily have to change. An observation made previously [11] highlights that one of the main differences in the results is the range of the evaluation metrics.

*4.1.6 Statistical Testing*

Finally, the last stage in the proposed protocol concerns the outcome of the evaluation step: any statistical testing procedure performed over such data should be accurately reported and, if possible, a motivation of its usage should be included. Check for significance in the results, or in the difference between a method and a baseline is possible via various means, e.g., paired/unpaired tests, effect size, confidence intervals, etc. [73]. This cements the importance of clarity and granularity of details while, additionally, to facilitate the interpretation of the results, related statistics such as the mean, variance, and population size of the samples should be reported.

More importantly, when performing any type of statistical testing, the data on which statistical method was computed must be specified. As with many aspects in recommendation, there is no standard procedure for this. This is especially problematic in cases when using cross-validated partitions, in turn using more than one test partition and, as an effect obtaining more than one performance measurement. A possibility due to this is drawing inconsistent conclusions about the significance of the results if performed on a split basis, e.g., a significant difference is found in some folds but not in others. On the other hand, concatenating the results from each split (on a user basis, as done in Information Retrieval for queries), may distort the test, due the data points not being independent (one user appears more than once) and the number of points increases substantially [15, 54].

4.2 Instantiations of Accountable Experimental Frameworks

In the previous section, we defined a set of requirements for reproducible recommender systems evaluation. Now, we provide an instantiation of such a framework, or, in other terms, how we implement the different stages involved in the recommendation pipeline represented in Figure 1.

For this, we need to be aware that most recommendation frameworks already define some evaluation functionality. Commonly, a number of evaluation metrics are available, together with some utility class allowing to execute recommendation and get the performance values corresponding to those metrics.

More specifically, these are the capabilities – at the moment of writing – of a selection of libraries in Recommender Systems research in terms of the stages previously introduced:

- Apache Mahout (version 14.1[20]): data splitting (on-the-fly), recommendation, evaluation (error and ranking metrics).
- CaseRec (version 1.1.0[21]): data splitting, recommendation, evaluation (error and ranking metrics), statistical testing.
- LensKit (version 3.0 in Java[22] and 0.10.1 in Python[23]): data splitting, recommendation, candidate item filtering (not in Python), evaluation (error and ranking metrics).
- LibRec (version 3.0.0[24]): data splitting, recommendation, evaluation (error and ranking metrics).
- MyMediaLite (version 3.12[25]): data splitting (on-the-fly), recommendation, evaluation (error and ranking metrics).
- RankSys (version 0.4.3[26]): recommendation, candidate item filtering, evaluation (ranking metrics).
- Surprise (version 1.1.1[27]): data splitting (on-the-fly), recommendation, evaluation (error metrics).

Since they are mostly focused on providing a (sometimes exhaustive) set of recommendation algorithms, none of these frameworks satisfies at the same time all the requirements presented before. Even when some requirements are fulfilled, they are not designed to support a reproducible and transparent environment. The most clear example can be found with the second stage – data splitting –, which in most situations is the first stage in the experiment (since the dataset is often created by an external party). For instance, RankSys does not perform any data splitting, while some frameworks (such as Mahout and MyMediaLite) create the splits in memory, generating different splits each time the experiment is run – unless some seed parameter is set. Nonetheless, there are other open source recommendation frameworks that support, to some extent, reproducible experimental settings:

- DaisyRec (unversioned[28]) is a recent framework developed in pyTorch focused on benchmarking previous research works, as described by Sun et al. [79]. It does not support statistical testing, error metrics, or the candidate item filtering component presented before, but it implements different strategies for negative sampling.

---

[20] https://github.com/apache/mahout/tree/mahout-14.1
[21] https://github.com/caserec/CaseRecommender/tree/1.1.0
[22] https://github.com/lenskit/lenskit/tree/lenskit-3.0-M2
[23] https://github.com/lenskit/lkpy/tree/0.10.1
[24] https://github.com/guoguibing/librec/tree/3.0.0/
[25] https://github.com/zenogantner/MyMediaLite/tree/v3.12
[26] https://github.com/RankSys/RankSys/tree/0.4.3
[27] https://github.com/NicolasHug/Surprise/tree/v1.1.1
[28] https://github.com/AmazingDD/daisyRec

- Idomaar (version 4.0.0[29]) was developed in the context of the EU FP7 project CrowdRec. It includes a suite of recommendation algorithms and supports both batch training and online streaming data based on Apache Kafka modules. It is focused on ranking metrics and allows to specify the entire computing environment as a virtual machine, providing an easy solution to repeat and reproduce a set of experiments. However, it does not support statistical testing.
- LibRec-auto (version 0.1.52[30]) is an extension of LibRec that allows to automate experiments by adding a layer on top of that library, although it may also work with custom recommenders. Because of this, it does not support the candidate item filtering component and is limited by the data splitting capabilities provided by LibRec; besides, it does not provide statistical testing. It handles the concept of *study*, which is a series of experiments executed with the same algorithm, dataset, and metrics. Being able to report these studies in configuration files should increase the reproducibility of the experiments performed with this framework. It has proved its flexibility in several works by e.g., Mansoury and Burke [61] and Sonboli et al. [77].
- RiVal (version 0.2[31]), presented in a previous work [70] as a framework oriented to the evaluation of *external* recommender systems. Its focus is on all the stages needed to evaluate an algorithm, not providing recommendations, instead it provides wrappers to recommendation libraries. This framework supports all the previously presented components. However, RiVal has not been in active development since 2015, thus the wrappers to external libraries are outdated. Apart from that, it sets a good example of how a reproducible framework might be defined. For instance, the experiments are defined through configuration files which can be shared easily, and each of the modules (split, recommend, evaluate) are independent, as a result, it can be used to run all the stages,l or a selection of them (for instance, with a given data split or a recommendation file generated in a different library).

Based on this, we observe that those frameworks which naturally fit in the recommendation pipeline depicted in Figure 1 make reproducibility easier and, as a consequence, the accountability of the generated experiments and of those research results obtained from them. In particular, we note that, to make those steps easier to integrate, each of the associated components should be able to communicate with the adjacent one. Many of the analyzed frameworks do this using plain text files, this allows making intermediate files publicly available, and, at the same time, facilitates integrating algorithms implemented in virtually any programming language, as long as the language supports file-based input and output.

In conclusion, we consider these toolkits very good approximations of a fair and transparent instantiation of the framework presented before. When used as a whole, they could improve the accountability of Recommender Systems research through the reproducibility of recommendation systems evaluation and their corresponding experimental settings.

---

[29] https://github.com/crowdrec/idomaar/tree/v4.0.0
[30] https://github.com/that-recsys-lab/librec-auto/tree/0.1.52
[31] https://github.com/recommenders/rival/tree/rival-0.2

**Table 3** Performance results in terms of NDCG using the framework-dependent evaluation.

| Algorithm | Framework | Lastfm | Movielens 1M | Yelp2013 |
|---|---|---|---|---|
| mf | LK | **0.806** | 0.954 | **0.986** |
|    | RS | NaN | 0.100 | 0.009 |
| ub | LK | 0.417 | **0.956** | 0.075 |
|    | RS | NaN | 0.060 | 0.008 |

**Table 4** Performance results in Yelp2013 using the controlled evaluation protocol.

| Algorithm | Framework | Candidate item | | | | | |
|---|---|---|---|---|---|---|---|
|   |   | TrainItems | | | UserTest | | |
|   |   | P@10 | R@10 | NDCG@10 | P@10 | R@10 | NDCG@10 |
| mf | LK | 0.000 | 0.000 | 0.000 | 0.086 | **0.979** | **0.896** |
|    | RS | 0.004 | 0.049 | 0.029 | 0.074 | 0.975 | 0.882 |
| ub | LK | 0.000 | 0.001 | 0.001 | **0.155** | 0.906 | 0.825 |
|    | RS | **0.006** | **0.060** | **0.036** | 0.082 | 0.784 | 0.709 |

## 5 Discussion, Challenges, and Limitations

The previous section proposed a reproducible framework for the evaluation of Recommender Systems, together with a review of available libraries and how well they could fit under this proposed architecture. However, it is not clear how comparable these libraries are with each other. This particular aspect was analyzed in our previous work [70] using the RiVal toolkit. That work, denoted the set of internal, available evaluation strategies accessible within a recommendation framework (such as Mahout) as a *framework-dependent evaluation protocol*, based on the default, internal protocols implemented in each framework. In general these strategies are not prone to provide accountable experimental environments, even though some sort of parameterization might be allowed. In contrast, the set of evaluation strategies that fulfill the requirements presented in this work were called *controlled evaluation protocol*, where one has full control of every step of the protocol; i.e., a reproducible/accountable experimental framework is used – in that case, RiVal. We now reproduce those experiments with a different set of datasets and libraries (more information about them is presented in the Appendix); the results are presented in Tables 3 and 4.

Based on the results shown in Table 3 a sharp difference between the two frameworks is noticeable in every dataset and with both recommendation algorithms. According to this, the most accurate framework of the two would be LensKit. One observation to note is that RankSys cannot produce a score for the Lastfm dataset. After investigating this, we identified this being due to how

NDCG is computed in RankSys, in particular, how the gains are generated. The library takes the rating that appears in the test set for a user-item pair (if it does not appear, then that item is not relevant), subtracts the threshold, and uses the result of this operation incremented in one as the power of 2 that will be considered for the gain of a particular item in a specific ranking position. However, this dataset does not contain ratings but user listenings, hence, in some cases these gains were $2^{99865} \neq 10^{33000}$, which is the reason for the undefined value returned by the RankSys evaluation protocol.

On the other hand, Table 4 shows the results for the controlled evaluation protocol, where the algorithms from RankSys (either mf or ub) always outperform other when using the TrainItems candidate item strategy, however, when using the UserTest strategy, LensKit obtains better results (we refer to the Appendix for the results using other datasets). These tables also show that, systematically, the performance values obtained with UserTest are higher than using TrainItems. This indicates that the UserTest candidate item strategy is modeling an easier (not realistic) task, where we know the user will rate something, and we want to optimize the ranking of those soon-to-be-rated items. Additionally, RankSys outperforms LensKit when using TrainItems. This result is only obvious in controlled evaluation, due to the framework-dependent evaluation protocol disguising the comparison by comparing incompatible evaluation protocols. Given these widely differing results, it seems pertinent that in order to realistically perform an inter-framework benchmarking, the evaluation process needs to be defined and both the recommendation and evaluation steps must be performed in a more controlled and transparent environment.

In the light of these results, it stands clear that even though different frameworks implement the recommendation algorithms in a similar fashion, **the results are not comparable**, i.e., the performance of an algorithm implemented in one framework cannot be compared to the performance of the same algorithm in another. Not only do there seem to exist differences in how the recommendation algorithms are implemented, but also how the evaluation methods themselves are brought to life. There is no de facto understanding on how (and on what) to evaluate a recommender algorithm. Coincidentally this also applies to how recommendation algorithms of a certain type should be realized (e.g., default parameter values and ad hoc implementations). This is not necessarily something negative. Yet, when it comes to performance comparison of recommender algorithms, a transparent and standardized (or controlled) evaluation is crucial. Without such, the relative performance of two or more algorithms evaluated under different conditions becomes largely meaningless. In order to objectively and definitively characterize the performance of an algorithm, a controlled evaluation, with a defined evaluation protocol is a prerequisite.

Additionally, given the large amount of recommender systems-related research being conducted, and the wide variety of established open source frameworks, as well as individually implemented code, it is necessary to not only describe in detail the algorithms used, but also the complete evaluation process, including strategies for data splitting, specifics on the evaluation metrics and methods, etc. This is consistent with other works in the area that also advocate for the use of guidelines to assist researchers [69]. Not only is this necessary in order to definitively show that one algorithm outperforms another – within the same framework –, but also to increase understanding of the algorithms (and their implementations), and allow for objective progress. It is important to note that inter-framework comparisons of recommendation quality can potentially point to incorrect results and conclusions, unless performed with great caution and in a controlled, framework independent, environment.

Simultaneously, there exist multiple other factors that make the evaluation of recommender systems very difficult, and at some points even flawed. Most recommender systems are used for the purpose of alleviating users in finding interesting information, this usually creates a bias in terms of popularity, i.e., the fact that some items are more popular than others creates synthetic similarities between users based solely on their interaction with popular items. If not taken into consideration, the evaluation of a recommender system can point to an algorithm performing significantly better than it would in a real world situation simply due to the fact that it will recommend mostly popular items [82], i.e., items that the user does not need to get recommended as there is a high probability that she already knows them. These biases that appear naturally create another challenge towards accountability and transparency, if, e.g., changes made to the dataset affecting these biases are not reported, or because a dataset might be more biased than another one [28], make it impossible to obtain the same results. The important issue here, would be that such biases (or other properties of the evaluation for that matter) are not well understood at the moment, and because of that, they are not properly acknowledged when designing the experiments nor when the settings are reported. In those cases, explanation-based approaches, such as Polatidis et al.'s [68], could alleviate some of these issues, if the researcher needs to be assisted about the experimental settings, and if we restrict ourselves to use one library.

It is also worth considering that, among the different requirements proposed in Section 4, dataset collection and statistical testing are heavily understated in present frameworks, and little research has been devoted to these aspects. More specifically, **reproducibility on how the datasets are created** is more difficult if not enough data are presented, such as the cleaning process, whether anonymization is applied, etc. (see [28]). On the other hand, **reproducibility in statistical testing** is necessary if we expect the reported results to be statistically significant, while ensuring some level of accountability and transparency in the process. For instance, there exist different ways to take samples for a test when doing cross-validation, in some cases, violating independence assumptions or evidencing flaws in the design [15]. As an example, in the Information Retrieval community there are studies about the extent of reporting statistical significance tests [20], however in Recommender Systems no such analyses have been made.

In summary, the issues of replication – obtaining the exact same results in the same setting – and reproducibility – obtaining comparable results using a different setting – are very difficult challenges at the moment, related to the issue of producing accountable recommender systems. They force researchers to reimplement the baseline algorithms to compare their approach against, or to pay extra attention to every algorithmic and evaluation detail, ignoring if the observed discrepancies with respect to what was already published come from omitted details from the original papers (parameters, methodologies, protocols, etc.) or a wrong interpretation of any of these intermediate steps. However, as already mentioned, this is an evolving scenario and has become a very active topic in Recommender Systems and related communities. In particular, there have been recent advances on how to improve reproducibility, by, e.g., Carterette and Sabhnani [21] and Di Buccio et al. [26]. In these works, different aspects are considered for this goal; for instance, Di Buccio et al. [26] propose to build a taxonomy of components of a (Information Retrieval) system in such a way that each component can be implemented as part of a pipeline; then, the performance of various pipelines can be measured and the effect of any of these components unequivocally gauged. Carterette and Sabhnani's work [21] provides a more theoretical solution by considering the raw outcomes of the experiments; observing that a low variance on the statistical analyses between systems is an indication of high reproducibility.

There are, nonetheless, other challenges raised by researchers from other communities. Bajpai et al. [7] shift the attention to the reviewing stage, where **double-blind review hinders reproducibility**. The work suggests, as a possible solution or potential motivation, to highlight reproducible papers, for instance, by spreading the use of badges such as the ones from ACM mentioned in Section 3.2. From a more practical perspective, Ferro [34] focuses on three aspects: the system runs (i.e., the execution of the algorithms), experimental collections (the actual dataset used, together with the ground truth data, and any other external information that may be needed), and meta-evaluation studies; the work considers these dimensions key to ensure reproducibility, but no perfect solution exists for any of them, because of that, it calls for an *orchestrated effort and a cultural change*. Under a more global, general viewpoint, Freire et al. [38] provide a list of open research challenges, including the fact that making a paper reproducible requires to change the behavior of scientists when doing research and communicating with others. Can this be done in an efficient way? Is there a long term impact for the researcher? If that was the case, the work argues that more researchers would be convinced on spending time making their work reproducible. Another challenge mentioned in this list is the need to **create research environments to enable reproducibility**. Related to the reproducibility as a process mentioned before, Lin and Zhang [60] argue that the resources used in a paper should be available (archived) for a long time, so we should provide infrastructures for this. Last, but not least, it is important to consider the cost of making something reproducible/replicable and of the associated documentation; in particular, the difficulties for newcomers should be taken into account if a community wants to lower the barrier and adopt general reproducibility guidelines.

To finish this section we want to highlight one important limitation of this work. In the context of Recommender Systems research, different stakeholders exist. In this work we only explore accountability for one of them: the consumer of the published research work. We have analyzed and instantiated several challenges, suggestions, and guidelines in the Recommender Systems research field, while **focusing on the reader of such research**. We consider the cases where the final user (i.e., the person using the recommender system) or the designer/developer are the ones requesting accountability outside of the context of this work. Those situations pose other problems that are different (although there might be some overlap) from those presented here. As discussed by Diakopoulos [27], when dealing with algorithmic decision making, every algorithm is different and must be understood in its context to determine what can be disclosed; this may be a technical process as well as a human-centered one. In our scenario, because of the very nature of the Recommender Systems field, such process will probably need both. Hence, as a community, we must develop a methodology of information-disclosure modeling that includes thinking about how the public would use any particular bit of information disclosed, combined with explicit strategies oriented to each of the stakeholders present in typical Recommender Systems.

## 6 Conclusions and Future Directions

Reproducibility in the Recommender Systems area is a topic that has gained traction in recent years. It has, de facto, been equated to the reproducibility of the algorithms involved in the experiments of the published papers, however, as we have presented, this is far from the truth. In this paper we address exactly this problem by providing a concrete definition of the problem and a proposal for a reproducible framework oriented to Recommender Systems research; as a consequence, we understand it would help to solve the issue for much of research published nowadays. For this, we have presented several lessons learned while developing recommendation algorithms and, specifically, when evaluating recommender systems. Based on these lessons, we have proposed a reproducible framework based on several stages that seeks, from the viewpoint of the recommender system practitioner or researcher, to rethink the whole recommendation pipeline in order to facilitate that works following this framework gain reproducibility and be easier to replicate. In particular, we have specified the following tasks in this pipeline: dataset collection, data splitting, recommendation, candidate item filtering, evaluation computation, and statistical testing. We also present and discuss a list of experimental libraries that, to some degree, fit such a framework.

We argue that, as a consequence, once we are capable of making reproducible research, we get accountable recommender systems. We do this by presenting an extensive survey of the concepts, not only limited to the Recommender Systems community, but related to the broader field of algorithmic accountability. As a positive aspect, we believe this equivalence should attract even more efforts from the community into improving reproducibility of the recommendation algorithms, as they will have a direct effect in producing more transparent and accountable Recommender Systems research. For instance, many of the current proposals contributed in the broader areas of computational research could (and should) be adapted and translated for recommendation [78].

However, the presented research has some limitations that should be addressed in the near future. We would like to understand the source of the discrepancies observed between internal and external evaluation results once a reproducible framework is used. On top of that, there is another problem that has not been addressed: how comparable are the different *reproducible* frameworks? While most of the works have focused on comparing published papers under a common framework, the use of more than one framework has not been considered, probably because the number of such frameworks is still very low.

Ideally, these frameworks should formalize a set of standardized evaluation methodologies, independent of specific recommendation algorithms or evaluation strategies. These methodologies shall remain transparent to the practitioners to help them address current policies on algorithmic accountability and transparency [52]. At the moment, theoretical works on specific components of the presented pipeline (i.e., the candidate item filtering one) have demonstrated that many works of research are not comparable, because the evaluation metrics computed on the strategies reported are not stable [57, 59]. This is a critical point deserving further attention, since it limits the validity of the results being published: even if we are able to perfectly reproduce the research work (because the author shared all the information, code, etc.) the results will be flawed.

Additionally, all the reproducible frameworks analyzed are completely focused on accuracy metrics. This is derived by the common scenario in the area where only those types of metrics are reported and optimized for. However, in the last years, beyond-accuracy metrics such as diversity, novelty, or serendipity are considered more, or at least as, important than accuracy. Because of that, in the future we aim to extend this proposal to other evaluation dimensions along with considering other contextual dimensions – like time and social networks – in the evaluation protocols, and also consider approaches to deal with statistical biases in the evaluation [12].

**Acknowledgements** This work has been funded by the Ministerio de Ciencia, Innovación y Universidades (reference: PID2019-108965GB-I00). The authors thank the reviewers for their thoughtful comments and suggestions.

**Table 5** Statistics of the three datasets used in the experiments.

| Dataset | Users | Items | Ratings | Density | Timespan |
|---|---:|---:|---:|---:|---|
| Lastfm | 1,892 | 17,632 | 92,834 | 0.28% | 2005-2011 |
| Movielens 1M | 6,040 | 3,706 | 1,000,000 | 4.47% | 2000-2003 |
| Yelp2013 | 45,981 | 11,537 | 229,907 | 0.04% | 2005-2013 |

## A Experimental settings

**Datasets.** In this work we use three datasets from different domains to evidence reproducibility problems that might arise when evaluating recommender systems. The three tested domains are movies (Movielens 1M[32]), music (Lastfm[33]), and user reviews of business venues (Yelp[34]). Two of these datasets contain users' preferences on items expressed as ratings on a 1-5 star scale (Movielens and Yelp), whereas one of them includes preferences as the (unbounded) number of interactions (artist listenings, in this case) between the users and the items. Even though not all the evaluated frameworks explicitly support implicit or log-based data such as the last one, we decided to include it as a further test on how these types of datasets are processed.

---

[32] Available at http://grouplens.org/datasets/movielens/

[33] Available at https://grouplens.org/datasets/hetrec-2011/

**Table 6** Performance results in Lastfm using the controlled evaluation protocol.

| Algorithm | Framework | Candidate item | | | | | |
|---|---|---|---|---|---|---|---|
| | | TrainItems | | | UserTest | | |
| | | P@10 | R@10 | NDCG@10 | P@10 | R@10 | NDCG@10 |
| mf | LK | 0.000 | 0.000 | 0.000 | 0.000 | 0.000 | 0.000 |
| | RS | 0.007 | 0.015 | 0.000 | 0.469 | **0.874** | **0.809** |
| ub | LK | 0.001 | 0.002 | 0.000 | **0.495** | 0.796 | 0.711 |
| | RS | **0.023** | **0.045** | **0.004** | 0.472 | 0.746 | 0.626 |

**Table 7** Performance results in Movielens 1M using the controlled evaluation protocol.

| Algorithm | Framework | Candidate item | | | | | |
|---|---|---|---|---|---|---|---|
| | | TrainItems | | | UserTest | | |
| | | P@10 | R@10 | NDCG@10 | P@10 | R@10 | NDCG@10 |
| mf | LK | 0.047 | 0.056 | 0.064 | 0.384 | 0.699 | 0.663 |
| | RS | **0.169** | **0.297** | **0.300** | 0.335 | 0.668 | 0.625 |
| ub | LK | 0.000 | 0.000 | 0.000 | **0.393** | **0.708** | **0.681** |
| | RS | 0.106 | 0.168 | 0.172 | 0.323 | 0.655 | 0.612 |

More specifically, the first dataset we use is Lastfm, which contains around 100,000 ratings from 1,892 users on 17,632 artists. The second dataset, Movielens 1M, with 1 million ratings from 6,040 users on 3,706 items. Finally, the Yelp2013 dataset, which apart from reviews also contains 229,907 ratings by 45,981 users on 11,537 businesses. The datasets additionally contain item meta data – e.g. movie titles, business descriptions, etc. – these data were not used in the scope of this work. A summary of their statistics is shown in Table 5.

**Recommendation frameworks.** In this paper we focus our analysis on two popular recommendation libraries still in active development: LensKit and RankSys. As stated in the introduction, both MyMediaLite and Mahout are not active at the moment and their use has decreased gradually. For a comparison between MyMediaLite, Mahout, and LensKit we refer the reader to Said and Bellogín [70].

**Algorithms.** In order to provide a comparison between the recommendation frameworks, we select two of the most common families of collaborative filtering algorithms: nearest neighbors and matrix factorization techniques. Specifically, we test a user-based nearest neighbor algorithm using a size of 50 for the neighborhood and the Cosine similarity (denoted as **ub**) and a matrix factorization technique with 50 factors (denoted as **mf**). For a comparison involving more algorithms, such as item-based nearest neighbor, see [70].

More specifically, we use the UserUserItemScorer recommendation algorithm with the CosineVectorSimilarity similarity in LensKit for ub, whereas in RankSys we use the UserNeighborhoodRecommender with VectorCosineUserSimilarity. For the mf algorithm, we use FunkSVDItemScorer in LensKit and MFRecommender with HKVFactorizer as factorizer in RankSys.

---

[34] Available at http://www.kaggle.com/c/yelp-recsys-2013/data

## B Complete results for controlled evaluation protocol

We configured RiVal to use a random partition where a global ratio of 80-20 is used to split the dataset into training and test; we decided to not check other splitting strategies based on our earlier findings [70] showing that the rest of strategies do not have a strong impact on the final performance. We used the algorithms described in the previous section and report RiVal's implementation of precision (P), recall (R), and normalized discounted cumulative gain (NDCG) at ranking position 10, with a relevance threshold of 5. The discounting function for NDCG is the exponential.

Tables 6 and 7 (together with Table 4, already presented in the main text of the manuscript) show the performance results in Lastfm, Movielens 1M, and Yelp respectively when the controlled evaluation is used. We observe an interesting pattern from these results: the algorithms from RankSys (either mf or ub) are always the best ones according to the TrainItems candidate item strategy, however, when using the UserTest strategy, LensKit tends to obtain better results (except for Lastfm dataset). This situation evidences one of the most critical aspects among the different stages in the proposed framework: the candidate item selection strategy affects heavily the final results, not only in terms of the metric value but also on the order of the recommendation algorithms. This result is in line with our previous work [70], and also previous research on evaluation methodologies [11].

These tables also show that, systematically, the performance values obtained with UserTest are higher than using TrainItems. This indeed indicates that the UserTest candidate item strategy is modeling an easier (not very realistic) task, where we know the user will rate something and we want to optimize the ranking of those soon-to-be-rated items. In fact, we observe that under these assumptions, the user-based algorithm is almost (and often better) as good as the mf, a counterintuitive result that does not match the general knowledge in the community.

A further observation to be made is how this controlled evaluation compares to each framework's internal evaluation protocol. By looking at the absolute numbers, we may infer that RankSys is internally using a candidate item selection very similar to the TrainItems, whereas LensKit is likely using UserTest or something very similar. This observation emphasizes the unfairness one may incur if we directly compare the values presented in Table 3, or, in general, if we want to compare two sets of results generated by different researchers or practitioners.